\documentclass[12pt,a4paper]{article}
\usepackage[english]{babel}
\usepackage[utf8]{inputenc}
\usepackage[dvips]{graphics}
\def\rfig#1#2{figure~#1\begin{figure}[htb]
{\it figure #1}\hfil\includegraphics{#2}\end{figure}}

\title{Using Virtual Addresses with Communication Channels}
\author{Oskar Schirmer}
\date{February 11th, 2013}
\begin{document}

\maketitle
\vfill

\section*{Abstract}

While for single processor and SMP machines,
memory is the allocatable quantity,
for machines made up of large amounts of parallel computing units,
each with its own local memory,
the allocatable quantity is a single computing unit.
Where virtual address management is used to
keep memory coherent and allow allocation
of more than physical memory is actually available,
virtual communication channel references
can be used to make computing units stay connected
across allocation and swapping.

\vfill
\newpage

\section*{Parallel Architecture}

For various reasons, an alternate design to SMP based
parallel computing for use with dynamic applications
is assumed to be implemented:
Large numbers of {\it computing units}, each composed of a
processing unit and local memory {\tt [1]}.
To allow {\it computing units} to cooperate, they shall be connected
by some network of comminucation channels.
Each {\it computing unit} being programmed much the same as
MMU-less micro controllers,
the full network is understood as a {\it parallel computing system} in the
sense of communicating sequential processes {\tt [2]}.
The single {\it computing units} should not differ in connectivity and
amount of local memory.

\section*{Resource Allocation and Usage}

On a single processor or SMP machine,
system global memory is the main resource to administer,
and it usually is portioned in memory pages
(\rfig{1}{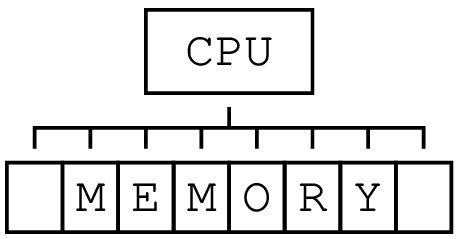}).

On a {\it parallel computing system} however,
the {\it computing units} are the main resource,
already portioned in units as is:
Running an application will make use of an arbitrary,
not necessarily fixed, number of {\it computing units}
(\rfig{2}{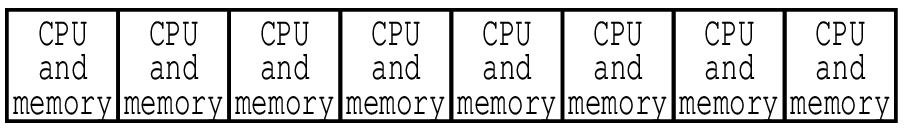}).

I.e., memory is never a passive resource to allocate,
but always served in conjunction with processing units.
It is not accessed thru memory addresses,
but thru communication channels, which in turn
are addressed using a number for each {\it channel end},
the {\it channel end address}.
For {\it computing unit} ``A'' to access some {\it computing unit} ``B'',
it configures
its {\it channel end} ``a'' to communicate to {\it channel end} ``b''
at {\it computing unit} ``B''
and subsequently transmits data over the channel.
The network layer of the channel is managed automatically
by some {\it interconnect node} hardware,
a good real example is provided by XMOS {\tt [3]}.

\section*{Resources Demand versus Availability}

To cope with the need of application programmes for
larger amounts of memory than actually are available
on a system, and to avoid fragmentation,
virtual address translation has been introduced
for single processor systems in 1977.\footnote{{\it VAX}
by {\it Digital Equipment Corporation}}
Through virtual address translation,
the main resource
used by some application (virtual memory) 
is mapped to the main resource
offered by the machine (physical memory, \rfig{3}{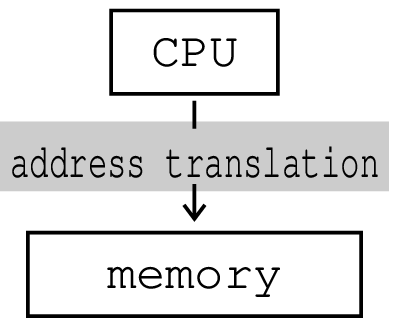}).

Now, with {\it computing units} being the main resource
on a {\it parallel computing system}, them being
referenced by the {\it channel end addresses} they offer,
we introduce a differentiation between
virtual {\it channel end addresses}, as used by applications,
and physical {\it channel end addresses}, as needed by the
{\it interconnect node} hardware.
This scheme asks for an equivalent to address translation
tables, some {\it channel end address} translation means.

\section*{Channel End Address Translation}

Comparing this {\it channel end address} translation to
conventional memory address translation,
{\it channel end address} translation tables have to be provided.
Different approaches lend itself to implement such tables:
\begin{itemize}
\item Explicit address translation is provided by some
dedicated {\it computing unit}
\item Implicit address translation is performed automatically
by some single facility at a single central location.
When establishing a connection,
the initiating {\it computing unit} automatically requests the
{\it channel end address} translation at the central facility
to find the physical destination
\item Implicit address translation is performed automatically,
but is distributed, e.g. equally to the {\it interconnect nodes}.
When establishing a connection,
the initiating {\it computing unit} ``A'' automatically requests the
{\it channel end address} translation
at the responsible {\it interconnect node}
(next to {\it computing unit} ``T'')
to find the physical destination, {\it computing unit} ``B''
(\rfig{4}{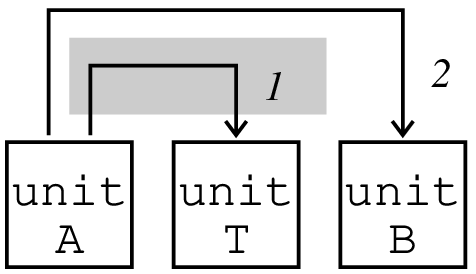})
\end{itemize}

In fact, there are real examples of similar systems.
One is the {\it domain name system} (DNS) that translates
node names into {\it IP addresses}.
Here, from the application point of view,
translation is accomplished explicitely.

\section*{Implementation}

With automatic route establishing already given,
and to avoid performance drop,
{\it channel end address} translation should be
implemented as an automatic feature, too.
To avoid congestion at a central facility,
and because virtual addresses may be chosen
without regard to the numbering,
address translation tables, and thus virtual addresses,
shall be provided per {\it interconnect node}.
E.g., the upper part of the {\it channel end address}
may be used to determine the {\it interconnect node} responsible,
the lower part of the address being the virtual address part,
which, by using it to index the translation table at
the {\it interconnect node} next to {\it computing unit} ``T'',
is translated to a full physical {\it channel end address}.
This physical {\it channel end address} is returned to the establishing
{\it computing unit} for further establishing the connection
to the physical destination, {\it computing unit} ``B''.

Whether it is favourable to keep translation results in
local caches for repeated use by the establishing {\it computing unit},
is subject to research. From the point of view of system
simplicity, caches should be avoided altogether.

{\it Channel end address} translation may fail.
On a system that supports exception handling,
an appropriate exception handler might be triggered
on {\it computing unit} ``A'',
or on {\it interconnect node} ``T''.
Again, for reasons of simplicity,
it may be desirable to avoid exceptions altogether.
To achieve this, the translation facility should
allow for configuration to send a message,
some {\it exception signal},
to a dedicated {\it computing unit},
which in turn is responsible for handling
the failure by loading or swapping appropriate code.

\section*{References}

\begin{description}
\item {\tt [1]} Eds. R. J. Elliott, C. A. R. Hoare:
{\it Scientific Applications of Multiprocessors},
1989, Prentice Hall
\item {\tt [2]} C. A. R. Hoare:
{\it Communicating Sequential Processes},
1985, Prentice Hall International
\item {\tt [3]} David May:
{\it The XMOS XS1 Architecture}, 2009, XMOS Ltd.
\end{description}

\end{document}